\documentclass[%
reprint,
amsmath,amssymb,
aps,
]{revtex4-2}

\usepackage{graphicx}
\usepackage{dcolumn}
\usepackage{bm}
\usepackage{braket}
\usepackage{mathtools}
\usepackage{bbold}
\usepackage{tikz}
\usepackage{soul}

\begin{document}


\title{Preventing the Breakdown of Tight-Binding Waveguide Optics by Löwdin Orthogonalization}


\author{Konrad Tschernig}
\email[]{konrad.tschernig@tu-berlin.de}

\affiliation{Deutsches Zentrum für Luft- und Raumfahrt e.V. (DLR), Institute of Optical Sensor Systems, Rutherfordstr. 2, Berlin 12489, Germany}
\affiliation{Technische Universit{\"a}t Berlin, Institut f{\"u}r Luft- und Raumfahrt (ILR), Marchstraße 12-14, Berlin 10587, Germany}

\author{Florian H. Huber}

\affiliation{Fakult{\"a}t f{\"u}r Physik, Ludwig-Maximilians-Universit{\"a}t M{\"u}nchen, Schellingstr. 4, 80799 M{\"u}nchen, Germany}
\affiliation{Max Planck Institute for Quantum Optics, Garching 85748, Germany}
\affiliation{Munich Center for Quantum Science and Technology (MCQST), Schellingstr. 4, 80799 M{\"u}nchen, Germany}
\affiliation{Department of Physics, Technische Universit{\"a}t M{\"u}nchen, James-Franck-Str. 1, 85748 Garching, Germany}

\author{Janik Wolters}

\affiliation{Deutsches Zentrum für Luft- und Raumfahrt e.V. (DLR), Institute of Optical Sensor Systems, Rutherfordstr. 2, Berlin 12489, Germany}
\affiliation{Technische Universit{\"a}t Berlin, Institut f{\"u}r Luft- und Raumfahrt (ILR), Marchstraße 12-14, Berlin 10587, Germany}

\author{Jasmin Meinecke}

\affiliation{Fakult{\"a}t f{\"u}r Physik, Ludwig-Maximilians-Universit{\"a}t M{\"u}nchen, Schellingstr. 4, 80799 M{\"u}nchen, Germany}
\affiliation{Max Planck Institute for Quantum Optics, Garching 85748, Germany}
\affiliation{Munich Center for Quantum Science and Technology (MCQST), Schellingstr. 4, 80799 M{\"u}nchen, Germany}
\affiliation{Department of Physics, Technische Universit{\"a}t M{\"u}nchen, James-Franck-Str. 1, 85748 Garching, Germany}



\date{\today}

\begin{abstract}
Many advancements in optics have relied on the tight-binding approximation, which simplifies the description and prediction of complex system behaviors. This approximation describes the dynamics of the total light field by examining the coupling between the guided modes of individual single-mode substructures -- also known as coupled mode theory. However, the underlying assumption, that the guided modes of individual waveguides form an orthogonal basis, breaks down when waveguides are brought into close proximity or when larger arrays are considered. In this work, we systematically analyze the consequences of this non-orthogonality and show that it leads to a generalized eigenvalue problem involving an overlap matrix, causing a fundamental mismatch between the standard TB model and solutions of the paraxial wave equation. To resolve this issue, we introduce a modified TB framework based on the Löwdin orthogonalization, which constructs an orthonormal basis from the non-orthogonal guided modes while minimally altering their physical shape and preserving their symmetry properties. The resulting Löwdin-TB method restores the standard eigenvalue problem and yields excellent agreement with exact beam propagation simulations across a wide range of system sizes and waveguide separations. Furthermore, it captures important physical effects, such as enhanced long-range coupling and nontrivial hopping phases, that are absent in the standard approach. 
\end{abstract}


\maketitle

\section{Introduction}
The tight-binding (TB) method, as a semi-empirical approach, was introduced in the field of solid-state physics, where it is also known as the linear combination of atomic orbitals (LCAO) method, and is primarily used to calculate the electronic band structure and single-particle Bloch states of materials \cite{slaterSimplifiedLCAOMethod1954}. More precisely, the TB method is used to simplify complex interactions within a crystal lattice by considering electrons as being tightly bound to atoms, with their wavefunctions overlapping only slightly with neighboring atoms. Over the years, researchers have recognized the TB method's wide applicability. It is now used in many areas of physics, chemistry, and even biology due to its computational efficiency and ability to provide intuitive insights into the properties of complex systems \cite{goringeTightbindingModellingMaterials1997}. This is because the TB method simplifies the analysis of the system by mapping a continuous variable description to a discrete, manageable set of parameters
and enables researchers to explore the effects of various perturbations, such as internal imperfections or external driving, with comparatively little effort. 
For example, nowadays, TB models are used in the study of phononic crystals \cite{mattarelliPhononicCrystalsSpherical2013}, Bose-Einstein condensates \cite{smerziNonlinearTightbindingApproximation2003,efremidisLatticeSolitonsBoseEinstein2003}, meta-materials \cite{xuTightbindingAnalysisCoupling2011}, molecular interactions \cite{sankeyInitioMulticenterTightbinding1989}, and protein folding \cite{mousaviProteinChainsTightbinding2025} to name a few. In optics and photonics, the TB method -- or coupled mode theory -- has proven to be a powerful tool for modeling and simulating the behavior of photonic crystals \cite{buttRecentAdvancesPhotonic2021,bayindirTightBindingDescriptionCoupled2000}, optical resonators \cite{fanTemporalCoupledmodeTheory2003}, optical cavities \cite{liCoupledModeTheory2010} and waveguide arrays \cite{szameitDiscreteOpticsFemtosecondlaserwritten2010,chenTightbindingModelOptical2021,amiriTightBindingAnalysisCoupled2006} among others. The latter field has spawned many significant advances in photonics, such as the observation of topological states \cite{kangTopologicalPhotonicStates2023,rechtsmanPhotonicFloquetTopological2013,harariTopologicalInsulatorLaser2018,bandresTopologicalInsulatorLaser2018}, Anderson localization \cite{segevAndersonLocalizationLight2013} or artificial gauge fields \cite{lumerLightGuidingArtificial2019}. Beyond classical light evolution, the TB method is crucial to model the behavior of few- or many-photon states in waveguide structures, leading to photonic quantum random walks \cite{pouliosQuantumWalksCorrelated2014} and applications in quantum information processing \cite{tschernigMultiphotonDiscreteFractional2018,tschernigTopologicalProtectionDegree2021}. \par
In this study we examine the limitations of the tight-binding method, specifically focusing on the propagation of light in arrays of waveguides. The fundamental equation that governs light propagation in these systems is the paraxial wave equation, which operates in an infinite-dimensional space of complex-valued square-integrable functions \cite{sroorModalDescriptionParaxial2021}. To simplify this vast solution space, the key concept of the tight-binding (TB) approximation is used, which describes the dynamics in terms of a finite-dimensional subspace formed by the guided modes of the individual waveguides. As a result, the overall dynamics can be represented as a finite-dimensional matrix-vector equation, allowing for efficient solutions on computers. However, a significant and often unacknowledged assumption behind this approach is the mutual orthogonality of the guided modes. Especially in the field of discrete quantum optics, the tight-binding Hamiltonian of the optical system under study often constitutes the starting point of any analysis of quantum light state evolution and the orthogonality of the underlying modes is simply taken for granted. In reality, this assumption is only approximately valid. For instance, the guided modes of two separate waveguides can be perfectly orthogonal only when the waveguides are infinitely far apart, a scenario where no light coupling can occur between them. As the distance between the waveguides decreases, the orthogonality assumption becomes increasingly violated. Ultimately, if the waveguides are too close, the tight-binding approximation fails, and its predictions diverge from the outcomes of the paraxial wave equation. In optics, this breakdown has been recognized before in the context of coupled mode theory \cite{huangCoupledmodeTheoryOptical1994,bellancaAssessmentOrthogonalNonorthogonal2018,schulzGeometricControlNextnearestneighbor2022}. However, a systematic and pedagogical discussion of the overlap problem in paraxial waveguide optics and its solution -- especially considering systems beyond 1D-waveguide arrays -- is missing in the literature and here we seek to fill this gap.  \par
We will first introduce the standard TB method in detail and with an emphasis on the role of the guided mode overlap. As we will show, the dominant reason for the breakdown of the standard TB method is the non-orthogonality of the guided mode basis which induces an often overlooked overlap matrix in the tight-binding equation of motion. As a direct consequence, the eigenvectors of the tight-binding Hamiltonian matrix do not correspond to eigenmodes of the principle (paraxial) Hamiltonian. Instead, the true eigenmodes (supermodes) of the paraxial Hamiltonian correspond to a generalized eigenvalue problem formed by the TB-Hamiltonian- and the overlap-matrix \cite{amiriTightBindingAnalysisCoupled2006}. The mismatch between the TB-Hamiltonian eigenvectors and the paraxial-Hamiltonian eigenmodes is then directly related to deviation of the overlap matrix from the identity matrix. Following this realization, we then introduce a straight-forward method to avoid the overlap-induced breakdown by orthogonalizing the guided mode basis using the Löwdin orthogonalization algorithm \cite{lowdinNonOrthogonalityProblemConnected1950}. This method was originally proposed by P.-O. Löwdin in 1950 in the context of theoretical chemistry to resolve the non-orthogonality problem in the calculation of the elastic constants of the alkali halides. In the context of optics it was rediscovered decades later under the name of non-orthogonal coupled mode theory \cite{huangCoupledmodeTheoryOptical1994,bellancaAssessmentOrthogonalNonorthogonal2018,schulzGeometricControlNextnearestneighbor2022}. The Löwdin-transformation preserves the relevant symmetries of the guided mode basis and achieves the orthogonalization involving the smallest possible modifications to the original modes. Recasting the tight-binding Hamiltonian in this new Löwdin-basis renders the overlap-matrix to be the identity matrix $\mathbb{1}$ and the generalized eigenvalue problem to be a standard one. As we will show, the modified TB method (Löwdin-TB method) leads to a closer match with the paraxial wave equation, especially when the waveguides are quite close and numerous. Finally, we will discuss the break-down of the Löwdin-TB method at even shorter distances, where the individual waveguides are no longer separate and instead form multi-mode structures. In this regime, the sub-space spanned by the guided mode basis can no longer support the multi-mode dynamics even after the application of the Löwdin transformation. \par
\section{Results}
We set the stage by considering the paraxial wave equation
\begin{equation}
	2ik n_0 \partial_z \psi = -\vec{\nabla}^2 \psi +V(x,y) \psi
	\label{eq:paraxial}
\end{equation}
where $\vec{\nabla}^2 = \partial_x^2+\partial_y^2$ is the Laplace operator in the transverse coordinates and $\psi=\psi(x,y,z)$ the slowly varying envelope of the electric field of a light wave traveling in $z$-direction with wave number $k=2\pi/\lambda$ in a dielectric medium with the optical potential $V(x,y)=-k^2(n^2(x,y)-n_0^2)$ and the corresponding refractive index distribution $n(x,y)$ and reference refractive index $n_0$. We call $\hat{H}=-\vec{\nabla}^2 +V(x,y)$ the paraxial Hamiltonian, in clear separation from the yet to be introduced tight-binding Hamiltonian $\bm{H}$. In the strictest sense $n_0$ is an effective refractive index $n_{eff}$ which is the average refractive index the electric field experiences. For an eigenmode of $\hat{H}$ with eigenvalue $\beta$, as derived later, the effective refractive index is given by $n_{eff}^2 = n_b^2-\frac{\beta}{k^2}$, where $n_b$ is the refractive index of the background medium. This has no effect on the eigenmodes, therefore it is convenient to set $n_0 = n_b$. For the sake of simplicity we assume to be in the weakly guiding regime, such that the electric field is transverse to the propagation direction and all polarization modes are linear polarized and uncoupled \cite{tanabeWeaklyGuidingApproximation2024}. 
\begin{figure*}[t]
\centering
\includegraphics[width=\linewidth]{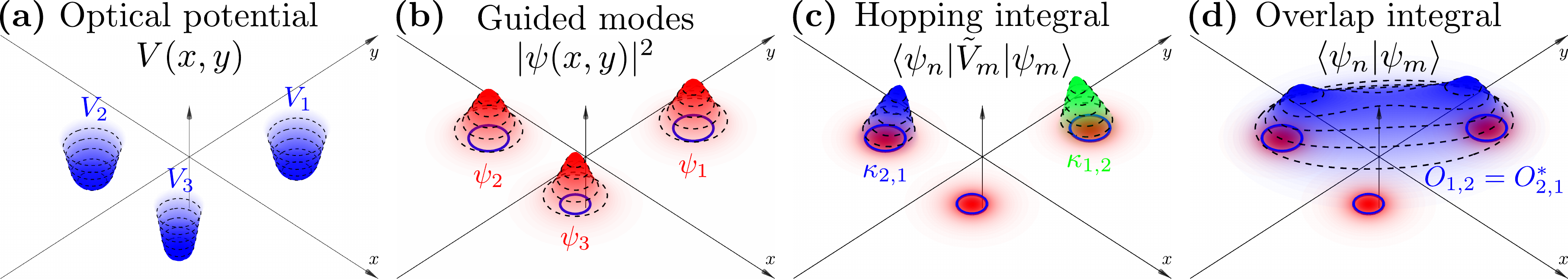}%
\caption{\label{fig:potential_guided_modes}\textbf{Standard Tight-binding Method} \textbf{(a)} Transverse optical potential in $(x,y)$-plane consisting of 3 waveguides (blue surfaces) with different shapes $f_m(x,y)$, refractive index contrasts $\Delta n_m$ and positions $(x_m,y_m)$. \textbf{(b)} Intensity distribution of the guided modes $|\psi_m(x,y)|^2$ (red surfaces) of the individual waveguides shown in (a) and their positions indicated by the blue circles. \textbf{(c)} The coupling coefficients computed from the hopping integrals $\kappa_{n,m}=\braket{\psi_n|\tilde{V}_m|\psi_m}=\int\psi^*_n(x,y)[V(x,y)-V_m(x,y)] \psi_m(x,y)~dx ~dy$, which is the volume under the green ($\kappa_{1,2}$) and blue surfaces ($\kappa_{2,1}$). \textbf{(d)} The overlap $\braket{\psi_1|\psi_2}=\int \psi^*_1(x,y) \psi_2(x,y)~dx~dy$ between two guided modes visualized as the volume under the blue surface. This quantity is assumed to be vanishingly small in the standard TB method. The dashed lines in (a-d) are surface contours to enhance the 3D visualization.}
\end{figure*}
A single waveguide is described by a refractive index profile of the form $n(x,y) = n_0+\Delta n f(x-x_0,y-y_0)$,  where $\Delta n$ is the refractive index contrast, $f(x,y)$ is the transverse refractive index profile and $(x_0,y_0)$ is the transverse position of the waveguide. In the weakly guiding regime the refractive index contrast of the waveguide is much smaller than the reference refractive index $n_0 \gg \Delta n$. This allows us to cast the optical potential, as shown in Fig.~(\ref{fig:potential_guided_modes}-a), in the form $V(x,y) = -2 k^2 n_0 \Delta n  f(x-x_0,y-y_0)$. Further, we assume that each waveguide supports only one guided mode, which is the solution to the eigenproblem
%
\begin{equation}
	\hat{H} \ket{\psi_0} = \beta_0 \ket{\psi_0},
	\label{eq:isolated_waveguides}
\end{equation}	
%
where $\beta_0$ is the propagation constant of the guided mode $\ket{\psi_0}=\psi_0(x,y)$, see Fig.~(\ref{fig:potential_guided_modes}-b). In order to facilitate the transition to the TB method, we have introduced Dirac notation with the associated inner product $\braket{\phi|\hat{A}|\psi} = \int \int \phi^*(x,y) \hat{A} \psi(x,y) \mbox{d}x \mbox{d}y$. We now transition from a single waveguide to an array of $M$ waveguides by simply expanding the optical potential as the direct sum $V(x,y) = \sum_{m=1}^M V_m(x,y)$, where $V_m(x,y) = -2 k^2  n_0 \Delta n_m  f_m(x-x_m,y-y_m)$ is the optical potential of the $m$'th waveguide. \par
The central idea of the TB method is that the guided modes $\ket{\psi_m}$ of the isolated waveguides form a natural basis that span the subspace in which the complete dynamics occur. We thus expand the solution to the paraxial wave equation in terms of the guided mode basis $\ket{\psi(z)} = \sum_{m=1}^M \alpha_m(z) \ket{\psi_m}$. In this subspace we obtain the vectorized paraxial wave equation
\begin{equation}
	2ik n_0 \bm{O} \partial_z \bm{\alpha}(z) = \bm{H} \bm{\alpha}(z),
	\label{eq:vectorized}
\end{equation}
where $\bm{H}$ is the tight-binding Hamiltonian with the elements $\bm{H}_{n,m} = \bra{\psi_n} \hat{H} \ket{\psi_m}$, $\bm{O}$ is the overlap matrix of the guided modes with the elements $\bm{O}_{n,m} = \braket{\psi_n|\psi_m}$ and $\bm{\alpha}(z)=(\alpha_1(z),\ldots,\alpha_N(z))^T$ is the vector containing the expansion amplitudes of the solution in the guided mode basis $\alpha_n(z) = \braket{\psi_n|\psi(z)}$. At this point, the paths will split and we will first walk along the direction of the standard treatment of Eq.~(\ref{eq:vectorized}) which will lead to the standard tight-binding Hamiltonian in section (\ref{sec:standard_tb}). We will return to Eq.~(\ref{eq:vectorized}) in section (\ref{sec:loewdin_tb}) to illustrate the Löwdin method. \par 
\subsection{Standard TB Method}
\label{sec:standard_tb}
In the standard TB method we assert the orthogonality of the guided modes and thus set $\bm{O}=\mathbb{1}$. As we have mentioned above, such an assertion is only true when the waveguides are infinitely far apart. However, this assumption allows us to further simplify the computation of the elements of $\bm{H}$. Using Eq.~(\ref{eq:isolated_waveguides}) and the orthogonality assumption $\bm{O}_{n,m}=\delta_{n,m}$ we obtain $\bm{H}_{n,m} = \beta_m \delta_{n,m} +  \kappa_{n,m}$ where $\kappa_{n,m}=\bra{\psi_n}\tilde{V}_m \ket{\psi_m}$ are the coupling coefficients obtained between waveguides $n$ and $m$. Here, $\tilde{V}_m =\sum_{k\neq m}V_k = \sum_{k=1}^M V_k - V_m$ is the total optical potential with the $m$'th waveguide removed. To be precise, the standard tight-binding coupling coefficient is obtained from the integral $\kappa_{n,m} = \int \int \psi_n^*(x,y) \left[V(x,y)-V_m(x,y) \right] \psi_m(x,y)  \mbox{d}x \mbox{d}y$, which is often referred to as the ''overlap``-integral in the literature \cite{liTightbindingPhotonics2025,chenTightbindingModelOptical2021}. To avoid confusion, we will call this expression the ''hopping`` integral, see Fig.~(\ref{fig:potential_guided_modes}-c), and reserve the term ''overlap`` integral for the bare guided-mode overlap $\bm{O}_{n,m} = \braket{\psi_n|\psi_m}$, see Fig.~(\ref{fig:potential_guided_modes}-d). The diagonal elements $\bm{H}_{n,n} = \beta_n+\kappa_{n,n}$ are simply the propagation constants of the isolated waveguides $\beta_n$, which are corrected by the self-coupling $\kappa_{n,n}$ due to the presence of neighboring waveguides. In summary, the standard tight-binding method maps the paraxial wave equation to the matrix-vector equation
\begin{equation}
		2ik n_0 \partial_z \bm{\alpha}(z) = \bm{H} \bm{\alpha}(z).
		\label{eq:standard_tb}
\end{equation}
Once the elements of $\bm{H}$ have been computed it is straight-forward to solve Eq.~(\ref{eq:standard_tb}) with the formal solution $\bm{\alpha}(z)=e^{-i\bm{H}z/(2k n_0)}\bm{\alpha}(0)$. To construct the propagation matrix $\bm{U}(z)=e^{-i\bm{H}z/(2k n_0)}$ the standard procedure is to find the eigenvectors $\bm{\gamma}^{(n)}$ and eigenvalues $\lambda^{(n)}$ of $\bm{H}\bm{\gamma}^{(n)} = \lambda^{(n)}\bm{\gamma}^{(n)}$. Then the matrix-exponential $e^{-i\bm{H}z/(2k n_0)}$ can be computed efficiently using the spectral decomposition leading to
\begin{equation}
	e^{-i\bm{H}z/(2k n_0)} = \bm{Q} e^{-i\bm{\Lambda} z/(2k n_0)} \bm{Q}^{-1},
\end{equation}
where we have defined the matrix $\bm{Q}$ that contains the eigenvectors $\bm{\gamma}^{(n)}$ as columns and the diagonal matrix $\bm{\Lambda}$ with the eigenvalues $\lambda^{(n)}$ as diagonal entries. Finally, we reconstruct the full wave function after the propagation distance $z$ from $\ket{\psi(z)}=\psi(x,y,z) = \sum_{m=1}^M \alpha_m(z) \psi_m(x,y)$. Here, $\bm{\alpha}(0)$ is the vector of the expansion coefficients $\alpha_m(0) = \braket{\psi_m|\psi(0)}$ of the initial wave function $\ket{\psi(0)}=\psi(x,y,0)$ in the guided mode basis. \par
\begin{figure*}[t]
\centering
\includegraphics[width=\linewidth]{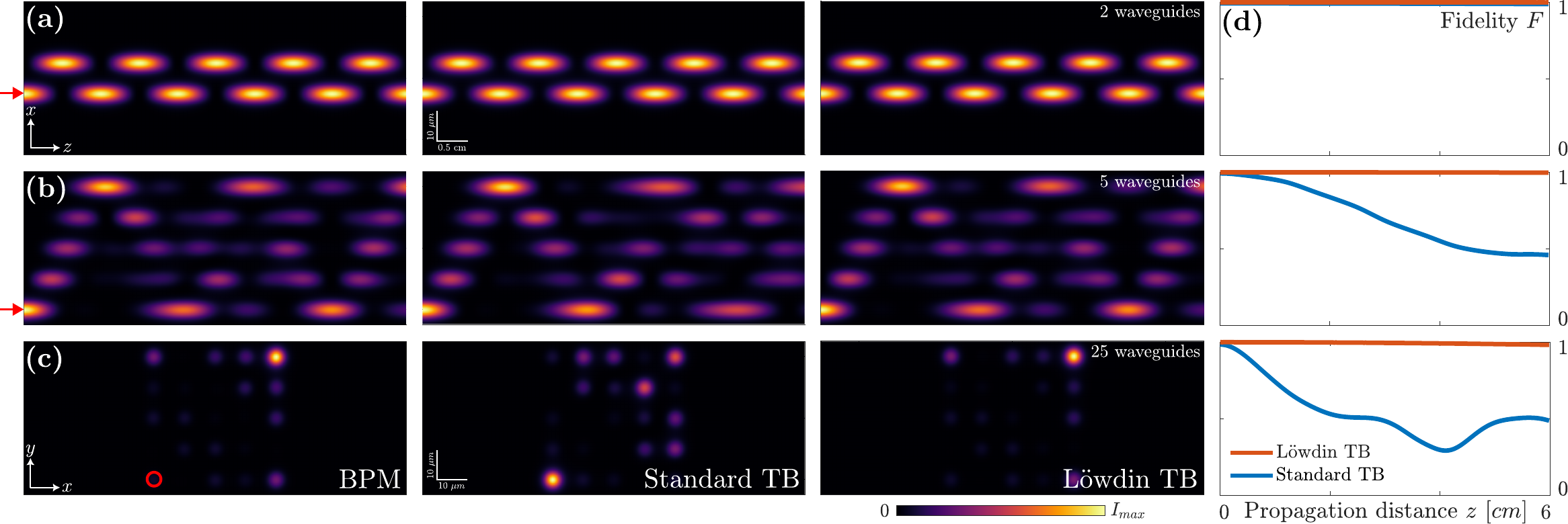}%
\caption{\label{fig:Standard_TB_fail}\textbf{Propagation in different Waveguide Systems} Intensity evolution in waveguide arrays with \textbf{(a)} $M=2$, \textbf{(b)} $M=5$ and \textbf{(c)} $M=25$ waveguides, respectively. The exact results are obtained by solving the paraxial wave equation using the beam propagation method (BPM), left panels, while the middle panels show the results of the standard tight-binding method and the right panels the Löwdin TB method. The red arrows (circle) indicate the input waveguide. In (a) and (b) the waveguides are arranged in a 1D lattice in the $x-z$-plane and we show the full evolution of the field in the $x-z$ plane (cut at $y=0$). In (c) the waveguides are arranged in a $5\times5$ square lattice and we show the field in the transverse $x-y$ plane at the final propagation distance $z=6$ cm. All waveguides are identical (width $\sigma = 3$~$\mu m$, contrast $\Delta n = 10^{-3}$, $n_0 = 1.45$, $\lambda=533$~$nm$, shape $f(x,y)=\exp(-|\vec{r}_\perp|^6/\sigma^6)$) and the lattice constant is $d=10$~$\mu m$ in all three cases. \textbf{(d)} Overlap (square root fidelity) of the TB- and BPM-solutions $F=|\braket{\psi_{BPM}|\psi_{TB}}|$.  While the standard TB method reproduces the BPM results faithfully for 2 waveguides, the fidelity degrades rapidly for a higher number of waveguides, which is also evident in the significantly different intensity patterns. The Löwdin TB method reproduces the BPM results with a fidelity above $99\%$ in all cases.}
\end{figure*}
As innocent as the last statement sounds, this is where we actually encounter the first hint of the breakdown of the standard tight-binding method. For instance, consider the initial wave function to be the guided mode of the $n$'th waveguide $\psi(x,y,0) = \psi_n(x,y)$. Intuitively, and in line with the orthogonality assumption, we then must have $\alpha_n(0) = 1$ and $\alpha_{m\neq n}(0) = 0$ otherwise, since we have only excited the $n$'th waveguide. In other words, we obtain the initial amplitude vector $\alpha_m(0) = \delta_{m,n}$. On the other hand, in any real waveguide array the waveguides must be at a finite distance from each other and thus $\braket{\psi_m|\psi_n}\neq \delta_{m,n}$. This leads to the initial vector $\alpha_m(0) = \braket{\psi_m|\psi(0)} = \braket{\psi_m|\psi_n} \neq \delta_{m,n}$; a clear contradiction to the intuitive result and it is not immediately clear which initial condition is the correct one. For the sake of consistency we will use the definition $\alpha_m(0) = \braket{\psi_m|\psi(0)}$ throughout this work, which takes the factual non-orthogonality of the guided modes into account, even though this choice contradicts the orthogonality assumption in the standard tight-binding approach. The fact that the initial excitation of a single guided mode unavoidably leads to a partial excitation of, in principle, all other guided modes is thus the first hint at the reason of the breakdown of the standard tight-binding method. We will now dive deeper to expose the full reason for the standard TB breakdown. 
\subsection{Breakdown of the Standard TB Method}
\label{sec:breakdown_standard}
In what follows we will show that the spectral decomposition of the standard tight-binding Hamiltonian differs from the spectral decomposition of the paraxial Hamiltonian when the guided modes are not an orthogonal set. This leads to a fundamental mismatch between the exact paraxial propagation operator and the TB-propagation operator and thus to the breakdown of the TB method. To see this, we begin with the paraxial eigenproblem 
\begin{equation}
	\hat{H} \ket{\chi^{(n)}} = \beta^{(n)} \ket{\chi^{(n)}},
	\label{eq:paraxial_eigen}
\end{equation}
where the wave functions $\ket{\chi^{(n)}}=\chi^{(n)}(x,y)$ must not be confused with the guided modes $\ket{\psi_m}=\psi_m(x,y)$ of the individual waveguides. Rather, the wave functions $\ket{\chi^{(n)}}$ are the proper eigen- or super-modes of the complete waveguide array with propagation constants $\beta^{(n)}$. By expanding the super-modes in the guided mode basis $\ket{\chi^{(n)}}=\sum_{m=1}^M \alpha^{(n)}_m \ket{\psi_m}$ we translate the paraxial eigenproblem into the guided mode subspace and obtain
\begin{equation}
	\bm{H} \bm{\alpha}^{(n)}= \beta^{(n)} \bm{O} \bm{\alpha}^{(n)},
	\label{eq:generalized_eigen}
\end{equation}
where $\bm{\alpha}^{(n)}$ is the vector of the expansion coefficients of the $n$'th super-mode in the guided mode basis. Crucially, this result shows that the standard eigenproblem of the paraxial wave equation, Eq.~(\ref{eq:paraxial_eigen}), is mapped to a generalized eigenproblem, Eq.~(\ref{eq:generalized_eigen}), in the guided mode basis, where the overlap matrix of the guided modes appears on the right-hand side \cite{amiriTightBindingAnalysisCoupled2006}.
\begin{figure*}[t]
\centering
\includegraphics[width=\linewidth]{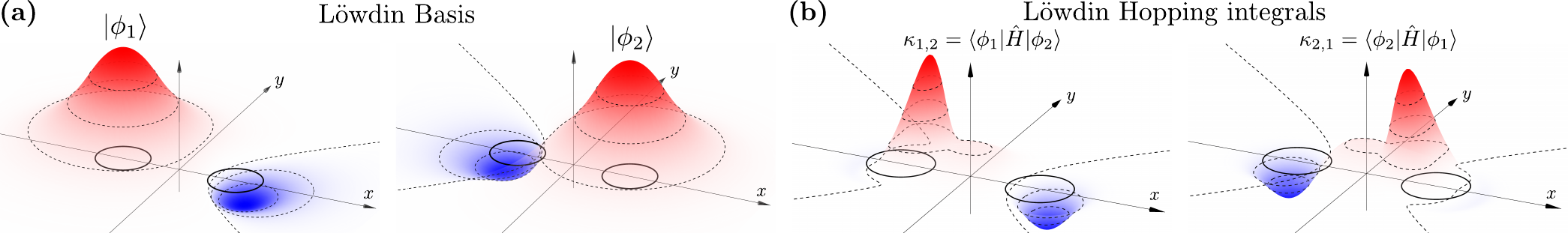}%
\caption{\label{fig:Löwdin_Tightbinding}\textbf{Löwdin Tight-binding Method} \textbf{(a)} Amplitude distributions $\ket{\phi_n}$ in the $x$-$y$-plane of the Löwdin modes for two identical coupled waveguides. Red (blue) surfaces correspond to positive (negative) amplitude. \textbf{(b)} Visualization of the hopping integrals in the Löwdin TB method. The Löwdin coupling coefficient $\kappa_{n,m} = \braket{\phi_n|\hat{H}|\phi_m}$ is the volume enclosed by the red (positive contribution) and blue (negative contribution) surface and the $x$-$y$ plane. In both (a) and (b) the position of the waveguides is indicated by the black circles and the dashed lines are surface contours to enhance the 3D visualization.}
\end{figure*}
Therefore, the eigenvectors $\bm{\gamma}^{(n)}$ and eigenvalues $\lambda^{(n)}$ of the standard TB Hamiltonian $\bm{H}$ can only coincide with the paraxial super-modes $\bm{\alpha}^{(n)}$ and propagation constants $\beta^{(n)}$ when the overlap matrix is the identity. This mismatch is highly nontrivial since the non-orthogonality changes the character of the eigenproblem and ultimately leads to the inevitable breakdown of the standard tight-binding method. In order to vividly showcase this breakdown we have simulated the light evolution in three different waveguide arrays, as shown in Fig.~(\ref{fig:Standard_TB_fail}). 
Unless stated otherwise we use the wavelength $\lambda = 533$~$nm$, reference refractive index $n_0 = 1.45$ and the waveguide shape function $f(x,y)=\exp(-|\vec{r}_\perp|^6/\sigma^6)$ with $\sigma = 3$~$\mu m$ and the refractive index contrast $\Delta n=10^{-3}$. These parameters lead to typical mode shapes encountered in femtosecond laser-written waveguides \cite{huberNonExponentialDecayFinite2025}. In the first case of two identical waveguides at a distance of $d=10$~$\mu m$, see Fig.~(\ref{fig:Standard_TB_fail}-a), we observe the well-known waveguide beam splitter evolution where the light hops back and forth between the waveguides \cite{tschernigMultiphotonDiscreteFractional2018}. 
As reference, we obtain the exact solution of the paraxial wave equation using the beam propagation method (BPM) \cite{pedrolaBeamPropagationMethod2015}. Here, the standard TB method reproduces the exact solution faithfully, which is evident in the close resemblance of the intensity distributions and the unit square root fidelity $F = |\braket{\psi_{BPM}(z)|\psi_{TB}(z)}|$ along propagation. This seems to imply that at this distance the guided mode overlap is sufficiently small and that the standard TB method is valid. However, this pictures changes drastically when we increase the number of waveguides to $M=5$ and $M=25$ in Fig.~(\ref{fig:Standard_TB_fail}-b,c). While the standard TB method initially remains close to the BPM solution, we see significant differences and a drop in fidelity to around 50\% further along the propagation. The most striking difference can be observed in Fig.~(\ref{fig:Standard_TB_fail}-c) where, according to the BPM, the initial waveguide (red circle) is almost completely dark after $z=6$~$cm$ propagation, while it contains the maximum intensity according to the standard TB method. To emphasize, the nearest-neighbor waveguide distance is $d=10$~$\mu m$ in all three cases, which highlights the non-trivial impact of the guided mode overlap matrix even if it is negligible for each pair of waveguides. \par
\begin{figure*}[t]
\centering
\includegraphics[width=\linewidth]{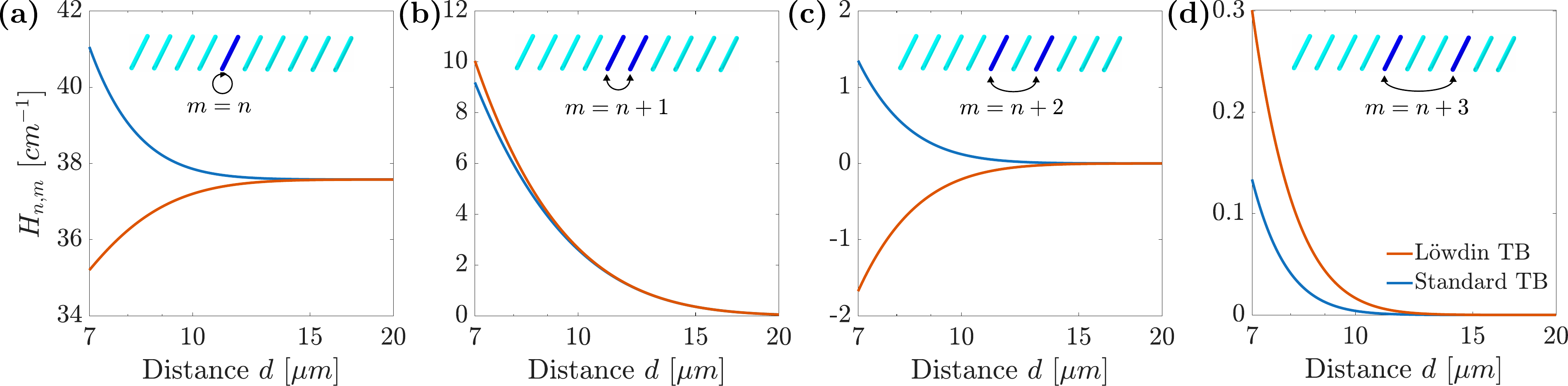}%
\caption{\label{fig:TB_coefficients}\textbf{Propagation Constants and Coupling Coefficients in Standard- and Löwdin-TB} We plot the matrix elements $\bm{H}_{n,m}$ of a linear array of 10 identical, coupled waveguides (see in-sets) as a function of the nearest neighbor distance $d$ for \textbf{(a)} $m=n$ (effective propagation constants), \textbf{(b)} $m=n+1$ (nearest neighbor coupling), \textbf{(c)} $m=n+2$ (next-nearest neighbor coupling) and \textbf{(d)} $m=n+3$ (next-next-nearest neighbor coupling). The blue (red) lines show the Standard-TB (Löwdin-TB) Hamiltonian. In the Löwdin-TB Hamiltonian the effective propagation constants are smaller and the coupling coefficients are larger than in the Standard-TB Hamiltonian. Importantly, the next-nearest neighbor coupling (when $m-n$ is even, as in (c)), $\bm{H}_{n,m}$ is negative, indicating a negative hopping phase that is unaccounted for in the Standard-TB method.}
\end{figure*}
\subsection{Löwdin TB Method}
\label{sec:loewdin_tb}
To prevent this breakdown we now return to Eq.~(\ref{eq:vectorized}) to remove the overlap matrix $\bm{O}$ in a more rigorous way. The essential idea is to expand the solutions of the paraxial wave equation not in the guided modes $\ket{\psi_n}$, but in a new set of modes $\ket{\phi_n}$ which form an orthogonal set. Then, by definition, $\bm{O}_{n,m}=\braket{\phi_n|\phi_m}=\delta_{n,m}$ and we obtain the standard form of the TB equation of motion, Eq. (4), except that now $\bm{H}_{n,m}=\braket{\phi_n|\hat{H}|\phi_m}$. Then the question arises as to how we choose the new set of modes $\ket{\phi_n}$? A popular choice are the super-modes $\ket{\phi_n}=\ket{\chi^{(n)}}$ of the paraxial Hamiltonian, as defined in Eq.~(\ref{eq:paraxial_eigen}), which is also known as the eigenmode expansion \cite{gallagherEigenmodeExpansionMethods2003}. While this choice yields the most accurate TB-approximation, the resulting TB-Hamiltonian is intrinsically diagonal. As a consequence there would be no coupling coefficients between the modes which complicates the use of such TB-Hamiltonians as coupled-system models. Moreover, the eigenmodes $\ket{\chi^{(n)}}$ are highly delocalized -- spanning over many waveguides -- in stark contrast to the guided modes that occupy mostly one waveguide. In addition, in the case of many waveguides, the super modes can be quite costly to compute, since the complete refractive index profile needs to be taken into account. Instead of the super-modes, we thus seek a set of orthogonal modes $\ket{\phi_n}$ which resemble the guided modes as closely as possible. To do so, we start with the guided modes $\ket{\psi_n}$ and remove the overlaps between them by applying a suitable orthogonalization procedure. The first idea that may come to mind is the well-known Gram-Schmidt algorithm \cite{leonGramSchmidtOrthogonalization1002013}. This algorithm converts a set of linearly independent vectors into an orthonormal basis that spans the same vector space, by iteratively subtracting projections of each vector onto the previously found orthonormal vectors and then normalizing the result. While this procedure yields a non-diagonal TB-Hamiltonian it features a significant draw-back due to the fact that the resulting shapes of the new modes $\ket{\phi_n}$ depend on the arbitrary enumeration of the guided modes. For example, the first guided mode $\ket{\psi_1}$ in the Gram-Schmidt process will remain unchanged $\ket{\phi_1} = \ket{\psi_1}$, while subsequent modes are increasingly distorted compared to the initial guided modes. Faced with the same conundrum, P.O. Löwdin introduced a new orthogonalization method which treats all starting vectors on an equal footing \cite{lowdinNonOrthogonalityProblemConnected1950}. Löwdin's procedure begins with the overlap matrix $\bm{O}$ and its spectral decomposition $\bm{O} = \bm{P} \bm{D} \bm{P}^\dagger$, where $\bm{P}$ is the unitary matrix containing the eigenvectors of $\bm{O}$ as columns and $\bm{D}$ is diagonal and contains the eigenvalues of $\bm{O}$. Using these we compute the inverse square-root $\bm{O}^{-1/2} = \bm{P} \bm{D}^{-1/2} \bm{P}^\dagger$ of the overlap matrix. The inverse square-root is guaranteed to exist, since the overlap matrix is a positive-definite matrix. Finally, $\bm{O}^{-1/2}$ contains the expansion coefficients of the new orthonormal basis in terms of the non-orthogonal guided modes
\begin{equation}
    \ket{\phi_n} = \sum_{m=1}^M \bm{O}^{-1/2}_{n,m} \ket{\psi_m}. \label{eq:Löwdin_Basis}
\end{equation}
This defines the Löwdin basis and it is easy to see that the $\ket{\phi_n}$ form an orthonormal set, since 
$\braket{\phi_k|\phi_l}=\sum_{n,m=1}^M  \bm{O}^{-1/2 *}_{k,n} \bm{O}_{n,m} \bm{O}^{-1/2}_{m,l} = \delta_{k,l}$. Löwdin showed that, among all possible choices of expansion matrices in Eq.~(\ref{eq:Löwdin_Basis}), $\bm{O}^{-1/2}$ yields the smallest modifications to the original guided modes \cite{lowdinNonOrthogonalityProblemConnected1950}. In this sense the Löwdin basis is unique, since it resembles the guided basis as close as possible while being an orthonormal set. Now that we have introduced the Löwdin-modes, we will discuss the differences and similarities with respect to the guided modes and the ramifications for the Löwdin-TB Hamiltonian. \par
In order to visualize the Löwdin basis, we consider the simplest case of $M=2$ coupled waveguides, with the overlap $\bm{O}_{2,1}^*=\bm{O}_{1,2} = \braket{\psi_1|\psi_2} = a<1 \in \mathbb{R}_+$ and $\bm{O}_{1,1} = \bm{O}_{2,2} = 1$. The inverse square root of $\bm{O}$ has the elements $\bm{O}^{-1/2}_{1,1}=\bm{O}^{-1/2}_{2,2}=\alpha$ and $\bm{O}^{-1/2}_{1,2}=\bm{O}^{-1/2}_{2,1}=\beta$. Using Eq.~(\ref{eq:Löwdin_Basis}) we find the Löwdin-Basis

\begin{align}
\begin{split}
    \ket{\phi_1} &= \alpha\ket{\psi_1}+\beta\ket{\psi_2}\\
    \ket{\phi_2} &= \beta \ket{\psi_1}+\alpha\ket{\psi_2},
\end{split}
\end{align}
where $\alpha = \frac{1}{2}(\sqrt{1+a}+\sqrt{1-a})/\sqrt{1-a^2}>1$ and $\beta = \frac{1}{2}(\sqrt{1-a}-\sqrt{1+a})/\sqrt{1-a^2}<0$. In general $|\alpha|>|\beta|$ and as a consequence, as shown in Fig.~(\ref{fig:Löwdin_Tightbinding}-a), the Löwdin-modes are mainly localized on their corresponding waveguide. However, depending on the magnitude of the guided mode overlap $a$, they are also localized on the opposite waveguide with a negative amplitude $\beta$. This extra support with a negative sign on the other waveguide guarantees the orthogonality of the Löwdin-modes, $\braket{\phi_1|\phi_2}=0$. Crucially, if $a \rightarrow 0$, i.e. when the waveguides are far enough apart so that the guided modes are effectively orthogonal, the Löwdin-modes become identical to the guided modes $\alpha \rightarrow 1$ and $\beta \rightarrow 0$ and we recover the Standard TB method. In other words, the Standard TB method is a special case of the Löwdin method when the waveguides are far away from each other. Once we have found the Löwdin-modes we can compute the Löwdin-TB Hamiltonian $\bm{H}$ and in Fig.~(\ref{fig:Löwdin_Tightbinding}-b) we show the Löwdin hopping integrals $\kappa_{1,2} = \kappa_{2,1}^* = \braket{\phi_1|\hat{H}|\phi_2}$. Notice, the integrands of the hopping integrals (blue and red surfaces) feature pronounced peaks which resemble the Standard TB-hopping integrals in Fig.~(\ref{fig:potential_guided_modes}-c) but also include non-trivial corrections due the orthogonality of the Löwdin-modes. \par 
\begin{figure*}[t]
\centering
\includegraphics[width=\linewidth]{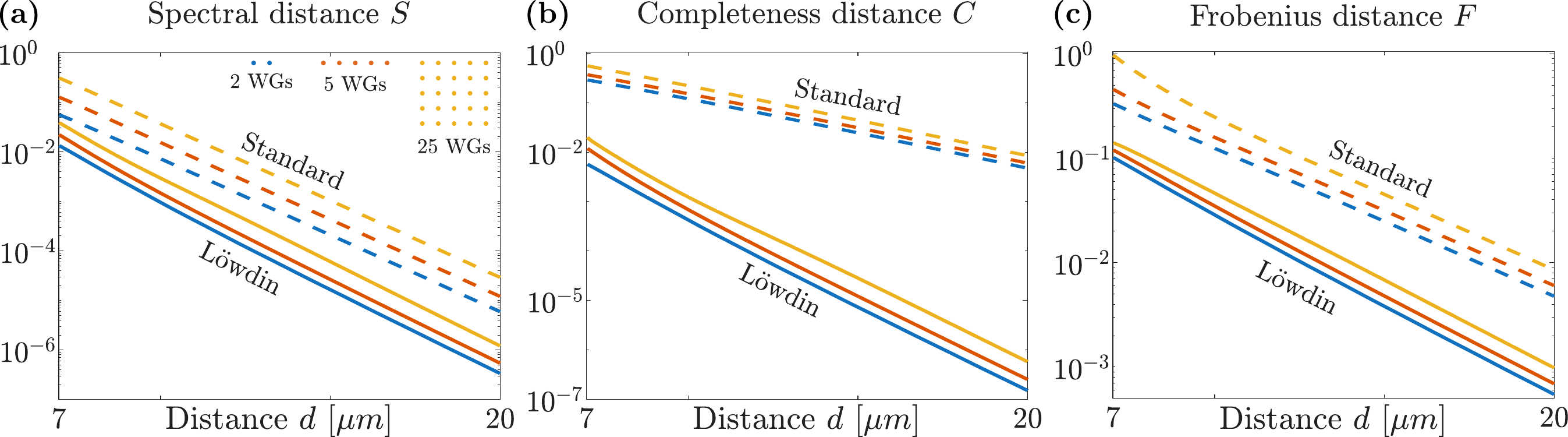}%
\caption{\label{fig:accuracy}\textbf{Comparison of the Accuracy of the Standard and Löwdin TB Methods} We plot the \textbf{(a)} Spectral distance $S$, \textbf{(b)} Completeness distance $C$ and \textbf{(c)} the Frobenius distance $F$ to the exact eigenmode expansion, according to Eqs.~(\ref{eq:spectral_distance}-\ref{eq:frobenius_distance}), of the Standard TB method (dashed lines) and Löwdin TB method (solid lines) as a function of the waveguide lattice constant $d$. The line colors correspond to 1D arrays with $M=2$ waveguides (blue) and $M=5$ waveguides (red) and a 2D array with $M=25$ waveguides (yellow). Smaller values of $S,C$ and $F$ correspond a better match with the exact eigenmode expansion. We observe a significant advantage for the Löwdin method over the whole range of distances $d$ in all three waveguide arrays and accuracy measures.}
\end{figure*}
As we have shown in Fig.~(\ref{fig:Standard_TB_fail}-a) both methods yield comparable results for a two-waveguide system. However, for larger 1D- and 2D- waveguide arrays, Fig.~(\ref{fig:Standard_TB_fail}-b,c,d), we observe a significant advantage for the Löwdin-TB method. Therefore, we now highlight the differences between the two methods for larger waveguide arrays in more detail. For this purpose we consider a linear array of 10 identical waveguides with the nearest neighbor distance (lattice constant) $d$. In Fig.~(\ref{fig:TB_coefficients}) we plot a selection of matrix elements of the Standard-TB and Löwdin-TB Hamiltonian as a function of the lattice constant $d$. Initially, we confirm that for larger distances, here $d>15~\mu m$, both methods yield practically identical results. However, once the guided mode overlap is non-negligible, $d<15~\mu m$, we observe significant differences. Firstly, see Fig.~(\ref{fig:TB_coefficients}-a), as $d$ decreases below $15~\mu m$, the diagonal elements $\bm{H}_{n,n}$ increase (decrease) for the Standard-TB (Löwdin-TB) method. In the case of the Löwdin method, this behavior is a result of the negative amplitude support of the Löwdin-modes on neighboring waveguides, which results in a negative correction with respect to the isolated waveguide propagation constants. The opposite happens in the Standard method, where the positive amplitude support of the guided modes on the neighboring waveguides results in a positive correction. Secondly, Fig.~(\ref{fig:TB_coefficients}-b), the nearest neighbor coupling behaves quite similarly in both methods and follows an exponential decay $\propto e^{-\xi d}$. Only below $d<10~\mu m$ are the coupling coefficients appreciably different, where the Löwdin-coupling is slightly stronger than the Standard-coupling. The most striking difference appears in the next-nearest neighbor (NNN) coupling coefficients, Fig.~(\ref{fig:TB_coefficients}-c), where the Löwdin coupling coefficients are strictly negative \cite{schulzGeometricControlNextnearestneighbor2022}. This surprising fact again originates from the negative amplitude components of the Löwdin modes rendering the contributions of the kinetic term negative and dominating over the positive optical potential term in the NNN coupling scenario. This results in an effective hopping phase of $\pi$ in the NNN Löwdin-coupling coefficients. In general we find that the condition for the negative hopping phase is fulfilled when $m-n$ is even, that is when there is an odd number of waveguides between the $n$'th and $m$'th waveguide. Such a mechanism is completely absent in the Standard TB method and highlights the importance of the waveguide overlaps in shaping the light dynamics. As in the nearest-neighbor case, we also observe that the absolute magnitude of the NNN coupling is slightly stronger in the Löwdin method. Lastly, in Fig.~(\ref{fig:TB_coefficients}-d), we show the 3rd neighbor coupling where we obtain the same positive coupling phase for both methods and a slightly stronger coupling for the Löwdin method. However, due to the 3 times increased distance to the 3rd neighbor the coupling is strongly suppressed compared to the nearest- and next-nearest neighbor coupling in both methods. In summary, the decrease in the effective propagation constants and the increase in the coupling coefficients lead to significantly enhanced long-range interactions between the Löwdin modes. Moreover, the effect of the negative NNN coupling phase is completely unaccounted for in the Standard TB method. All these discrepancies lead to the breakdown of the Standard TB method -- not only in the small distance regime, but also in the multi-waveguide case. \par
Finally, we compare the accuracy of the two methods more rigorously. As our measures of fidelity, we define the spectral-, completeness- and Frobenius distance of the respective TB Hamiltonians to the exact eigenmode expansion. The spectral distance is 
\begin{equation}
    S = \sqrt{\sum_{n=1}^M |\lambda_{TB}^{(n)}-\beta^{(n)}|^2/\sum_{m=1}^M |\beta^{(m)}|^2},
    \label{eq:spectral_distance}
\end{equation}
which is simply the mean, relative error of the TB eigenvalues $\lambda_{TB}^{(n)}$ of $\bm{H}$ with respect to the exact paraxial eigenvalues $\beta^{(n)}$. Further, we introduce the completeness distance
\begin{equation}
    C = \left|\left|\sum_{n=1}^M \ket{\chi_{TB}^{(n)}}\bra{\chi_{TB}^{(n)}}-\mathbb{1}\right|\right|/\left|\left|\mathbb{1}\right|\right|,
    \label{eq:completeness_distance}
\end{equation}
where $\ket{\chi_{TB}^{(n)}}=\sum_{m=1}^M \gamma^{(n)}_m \ket{\psi_m}$, $\bm{\gamma}^{(n)}$ is the $n$'th eigenvector of $\bm{H}$ and $\ket{\psi_m}$ are either the standard guided modes or the Löwdin modes according to the TB method used. We use the Frobenius matrix norm evaluated in the exact eigenmode basis $||\cdot||^2 = \sum_{n,m=1}^M |\braket{\chi^{(n)}|\cdot|\chi^{(m)}}|^2$. In other words, $C$ measures how far away the TB eigenmodes $\ket{\chi_{TB}^{(n)}}$ are from being a complete basis in the space of the exact paraxial eigenmodes. Lastly, we define the Frobenius distance
\begin{equation}
    F = \left|\left|\hat{H}_{TB}-\hat{H}_E\right|\right|/\left|\left|\hat{H}_E\right|\right|,
    \label{eq:frobenius_distance}
\end{equation}
which measures the combined effect of the spectral- and completeness errors. Here, $\hat{H}_{TB}$ is the projection of the TB Hamiltonian $\bm{H}$ into continuous space defined by the outer-product $\hat{H}_{TB}=\sum_{n=1}^M \lambda_{TB}^{(n)}\ket{\chi_{TB}^{(n)}}\bra{\chi_{TB}^{(n)}}$ and $\hat{H}_E = \sum_{n=1}^M \beta^{(n)}\ket{\chi^{(n)}}\bra{\chi^{(n)}}$ is the exact spectral decomposition of the paraxial Hamiltonian. In Fig.~(\ref{fig:accuracy}) we plot $S,C$ and $F$ as a function of the lattice constant $d$ in three different waveguide arrangements. As examples we choose the same waveguide arrays we used in Fig.~(\ref{fig:Standard_TB_fail}). We use a logarithmic scale for the vertical axis, which renders the exponential relationships to appear as linear. In all cases we observe a clear advantage for the Löwdin method (solid lines) over the Standard method (dashed lines). In the case of the spectral- and Frobenius distance, see Fig.~(\ref{fig:accuracy}-a,c) both the Standard and Löwdin method feature a similar scaling law, indicated by their almost parallel lines. This means that, in these measures, the Löwdin method has an approximately constant advantage of a factor $\approx10$ over the Standard method, regardless of the distance $d$. The most striking difference is observed in the completeness distance $C$, see Fig.~(\ref{fig:accuracy}-b), where the lines corresponding to the Löwdin method are significantly steeper. This implies that the advantage for the Löwdin method, in this measure, actually grows with the distance $d$, from a factor of $\approx 10^2$ ($d=7~\mu m$) to $\approx 10^5$ ($d=20~\mu m$). This corroborates our analysis of the breakdown of the Standard TB method due to the overlap-induced eigenmode mismatch. However, it is clear that both methods become increasingly inaccurate at smaller distances. The reason is because, as the waveguides get closer and closer, they form multi-mode structures and the exact paraxial eigenmodes become increasingly distorted with respect to the isolated guided modes. And, since the Löwdin modes are linear combinations of the guided modes, these non-linear effects can not be fully compensated by the Löwdin orthogonalization. Nevertheless, in light of these results, the Löwdin method constitutes a significant improvement over the Standard TB method.
\section{Summary \& outlook}

To summarize, we have investigated the limitations of the standard tight-binding method when applied to light propagation in waveguide arrays. While the Standard TB method is widely valued for reducing the infinite-dimensional paraxial wave equation to a tractable finite-dimensional system, it relies on a critical assumption: that the guided modes of individual waveguides are mutually orthogonal. We demonstrate that this assumption is only approximately valid and breaks down as waveguides are brought closer together, where mode overlap becomes non-negligible. This non-orthogonality introduces an overlap matrix into the formulation, which is simply neglected in the standard approach. As a result, the eigenvectors of the standard TB Hamiltonian no longer correspond to the true eigenmodes (supermodes) of the system, leading to discrepancies between TB predictions and exact solutions of the paraxial wave equation. These discrepancies become especially pronounced in larger waveguide arrays, even when pairwise overlaps appear small. To address this issue, we introduced an improved approach based on the Löwdin orthogonalization algorithm. This method constructs an orthonormal basis from the original guided modes while minimally altering their structure and preserving system symmetries. In this new basis, the overlap matrix becomes the identity, yielding a modified TB Hamiltonian (the Löwdin-TB method). Our numerical simulations show that this approach significantly improves agreement with exact solutions across a range of system sizes and waveguide separations. We further analyzed how the Löwdin-TB method modifies effective propagation constants and coupling coefficients, revealing enhanced long-range interactions and confirm previously encountered effects such as negative next-nearest-neighbor coupling phases \cite{schulzGeometricControlNextnearestneighbor2022}. Our quantitative error metrics confirm that the Löwdin method consistently outperforms the standard TB approach. However, both methods ultimately fail at very small waveguide separations (strong coupling), where the system transitions to a multi-mode regime beyond the scope of single-mode TB approximations. \par
Looking forward, this work opens several concrete directions for extending both the theoretical framework and practical applicability of tight-binding models in photonics and beyond. A natural next step is the application of the Löwdin-TB framework to non-Hermitian photonic systems \cite{zeunerObservationTopologicalTransition2015}. Since the method preserves Hermiticity while accurately incorporating overlap effects, it provides a more physically consistent platform for studying phenomena such as non-reciprocal transport. In particular, revisiting previously reported non-Hermitian effects in waveguide arrays through the lens of overlap-corrected models may lead to revised interpretations and new design principles  \cite{vicencioNonsymmetricEvanescentCoupling2025}. Furthermore, a generalization of the theory to bent or helical waveguides would be of high interest for the study of topological photonic lattices. In such systems the overlap- and hopping-matrices are not constant during propagation and their $z$-dependence non-trivially impacts the dynamics. From an engineering perspective, integrating the Löwdin orthogonalization into photonic design methods could open up the design space to stronger coupling regimes and thus smaller device footprints. This is especially relevant for large-scale photonic circuits and quantum optical networks, where full continuous space simulations are costly and propagation losses need to be avoided. Experimentally, the predictions of the Löwdin-TB model, such as modified propagation constants, coupling strengths and negative hopping phases, invite direct validation in engineered waveguide arrays over broad parameter spaces and beyond linear 1D arrays. Modern fabrication platforms (e.g., femtosecond laser-written waveguides) are well-suited to probe these effects. In summary, the Löwdin-TB approach provides a robust foundation for moving beyond idealized overlap assumptions, and its further development could play a key role in accurately modeling and designing complex waveguide systems.
\bibliography{tight-binding}

\end{document}